\documentclass[prd,superscriptaddress,floatfix,notitlepage,nofootinbib,reprint]{revtex4-1}  

\usepackage{graphicx}  
\usepackage{dcolumn}   
\usepackage{bm}      
\usepackage{amsmath,amssymb}
\usepackage{aas_macros}
\usepackage{multirow}
\usepackage{color}
\usepackage{verbatim}
\usepackage{tabu}
\usepackage{hyperref}
\usepackage{url}
\usepackage{microtype}
\usepackage[capitalise]{cleveref}
\usepackage[all]{hypcap}
\usepackage[toc,page]{appendix}
\usepackage{epstopdf}
\newcommand{\mr}{\mathrm}
\begin{document}
\widetext

\title{The Cross-correlation of KSZ Effect and 21 cm Intensity Mapping with Tidal Reconstruction}

\author{Dongzi Li}
\affiliation{Canadian Institute for Theoretical Astrophysics, University of Toronto, 60 St. George Street, Toronto, Ontario M5S 3H8, Canada}
\affiliation{Department of Physics, University of Toronto, 60 St. George Street, Toronto, ON M5S 1A7, Canada}
\affiliation{Dunlap Institute for Astronomy and Astrophysics, University of Toronto, 50 St. George Street, Toronto, Ontario M5S 3H4, Canada}

\author{Hong-Ming Zhu} 
\affiliation{Berkeley Center for Cosmological Physics and Department of Physics,
University of California, Berkeley, California 94720, USA}
\affiliation{Lawrence Berkeley National Laboratory, 1 Cyclotron Road, Berkeley, California 94720, USA}

\author{Ue-Li Pen} 
\affiliation{Canadian Institute for Theoretical Astrophysics, University of Toronto, 60 St. George Street, Toronto, Ontario M5S 3H8, Canada}
\affiliation{Dunlap Institute for Astronomy and Astrophysics, University of Toronto, 50 St. George Street, Toronto, Ontario M5S 3H4, Canada}
\affiliation{Canadian Institute for Advanced Research, CIFAR Program in Gravitation and Cosmology, Toronto, Ontario M5G 1Z8, Canada}
\affiliation{Perimeter Institute for Theoretical Physics, 31 Caroline St. N., Waterloo, Ontario, N2L 2Y5, Canada}

\date{\today}

\begin{abstract}
 We discuss the possibility of studying diffuse baryon distributions with  kinematic Sunyaev-Zel'dovich (kSZ) effect by correlating cosmic microwave background (CMB) temperature fluctuations with density fluctuations from 21\,cm intensity mapping (IM). 
The biggest challenge for the cross-correlation is the loss of large-scale information in IM, due to foregrounds and the zero spacing problem of interferometers. 
We apply the tidal reconstruction algorithm to restore the lost large-scale modes, 
which increases the correlation by more than a factor of three. 
With the predicted foreground level, we expect a $\sim20\,\sigma$ detection of kSZ signal for $0.8\lesssim z\lesssim2.5$ with CHIME and Planck, 
and a $\sim40\, \sigma$ detection with HIRAX and Planck. 
The significance can be greatly increased with next-generation facilities of higher spatial resolutions. 
\end{abstract}
\maketitle

\section{Introduction}
For $z\lesssim 2$, a large fraction of the predicted baryon content is missing in observations. The majority of these baryons are believed to reside in the warm-hot intergalactic medium (WHIM), with typical temperatures of $10^5$ K to $10^7$ K \cite{01Dave,Pen1999,Soltan06}. The low density imposes difficulties for direct detection. 
The uncertainty in the spatial distribution of its ionization state, metallicity, and pressure leads to confusion in interpreting signals from absorption lines and soft X-rays. 

The kinematic Sunyaev-Zel'dovich (kSZ) effect \cite{Sunyaev72,Sunyaev80,Vishniac87} is a promising probe for diffuse baryon content. 
KSZ signal is a secondary anisotropy in cosmic microwave background (CMB) temperature, which comes from the Doppler shift of photons induced by the radial velocity of free electrons. 
It has the following advantages: First of all, it receives a contribution from all the free electrons, 
which traces $\gtrsim 90\%$\,\cite{Fukugita04} the baryons at low redshifts. 
Second, the signal is mainly influenced by the electron density and the radial velocity, regardless of the temperature, pressure, and metallicity. Therefore, no extra assumptions are required to estimate the baryon abundance. Lastly, the radial velocity is a large-scale field, so the signal is less biased by the local environment, and more indicative of the diffuse distribution.
Hence, the kSZ is an unbiased probe for density fluctuations and its strength at different angular scales can be model-independently translated into baryon contents and diffuseness. 

Studying the kSZ effect is challenging as its relatively weak compared to the various contaminations, such as the primary CMB, thermal SZ effects, CMB lensing, and instrumentation noises. Another consideration is kSZ is a projected signal with contributions from different redshift mixed together. One way to mitigate the problem is, to cross-correlate CMB map with the density fluctuations from another tracer at a specific redshift.    
Several types of surveys have been proposed to play the role\,\cite{Hand12,Shao11,Li14,Hill16,Ferraro16}. 
Galaxy spectroscopic surveys, with accurate redshift information and high angular resolution, are powerful probes of density fluctuations at low redshift for high angular scales, i.e. $\ell>4000$ \cite{16Schaan}. However, the survey speed and cost limit their sky coverage and depth. Especially for $z\sim 1.4-2.5$, lack of spectral lines will lead to large shot noise in the measured density field. 
Projected field surveys, such as galaxy photometric surveys and gravitational lensing maps, on the other hand, can provide a densely sampled sky up to the high redshift. However, a significant fraction of kSZ signals come from the density fluctuations along the line of sight (LOS) due to the coupling of two fields (see section \ref{sec:kszAnalysis}). Cross-correlating it with a survey without LOS structure will inevitably lead to suboptimal correlation and loss of information.  

In this paper, we discuss the possibility of cross-correlating neutral hydrogen (HI) density field from 21 cm intensity mapping (IM) experiments with the kSZ signal.  21\,cm spectral lines provide accurate redshift information, and IM experiments perform fast scans of large sky area by  integrating all photons detected. In the following few years, there will be several large sky IM surveys producing data up to redshift 2.5 
\cite{2014CHIME, TIANLAI, HIRAX}. 
The foreground contaminations, zero-spacing of interferometers and small scale noises of IM are main factors that will downgrade the correlation.
The loss of large-scale information makes it almost impossible to cross correlate the IM surveys with other projected field
surveys.
After estimating the influence of these aspects for the correlation with kSZ effect, we demonstrate how the tidal reconstruction algorithm\cite{2012:pen,2015:zhu,2016Zhu,2018Foreman} can increase the correlation.

The paper is organized as follows: Section \ref{sec:kszRecon} describes how to cross-correlate density fields with CMB temperature fluctuations (following Ref. \cite{Shao11}); Section \ref{sec:21cm} addresses the limits of 21cm IM surveys and its influence on the correlations; 
 Section \ref{sec:kszAnalysis} demonstrates the scales of density and velocity fluctuations that contributes most in kSZ distortions; Section \ref{sec:tide} summarizes the tidal reconstruction algorithm which reconstructs the missing large-scale modes; Section \ref{sec:simu} presents numerical results and expected S/N; We conclude in Sec. \ref{sec:conclusion}.

\begin{figure}[tbp]
\begin{minipage}[t]{\linewidth}
\vspace{-0.6cm}
\center
\includegraphics[width=0.9\textwidth]{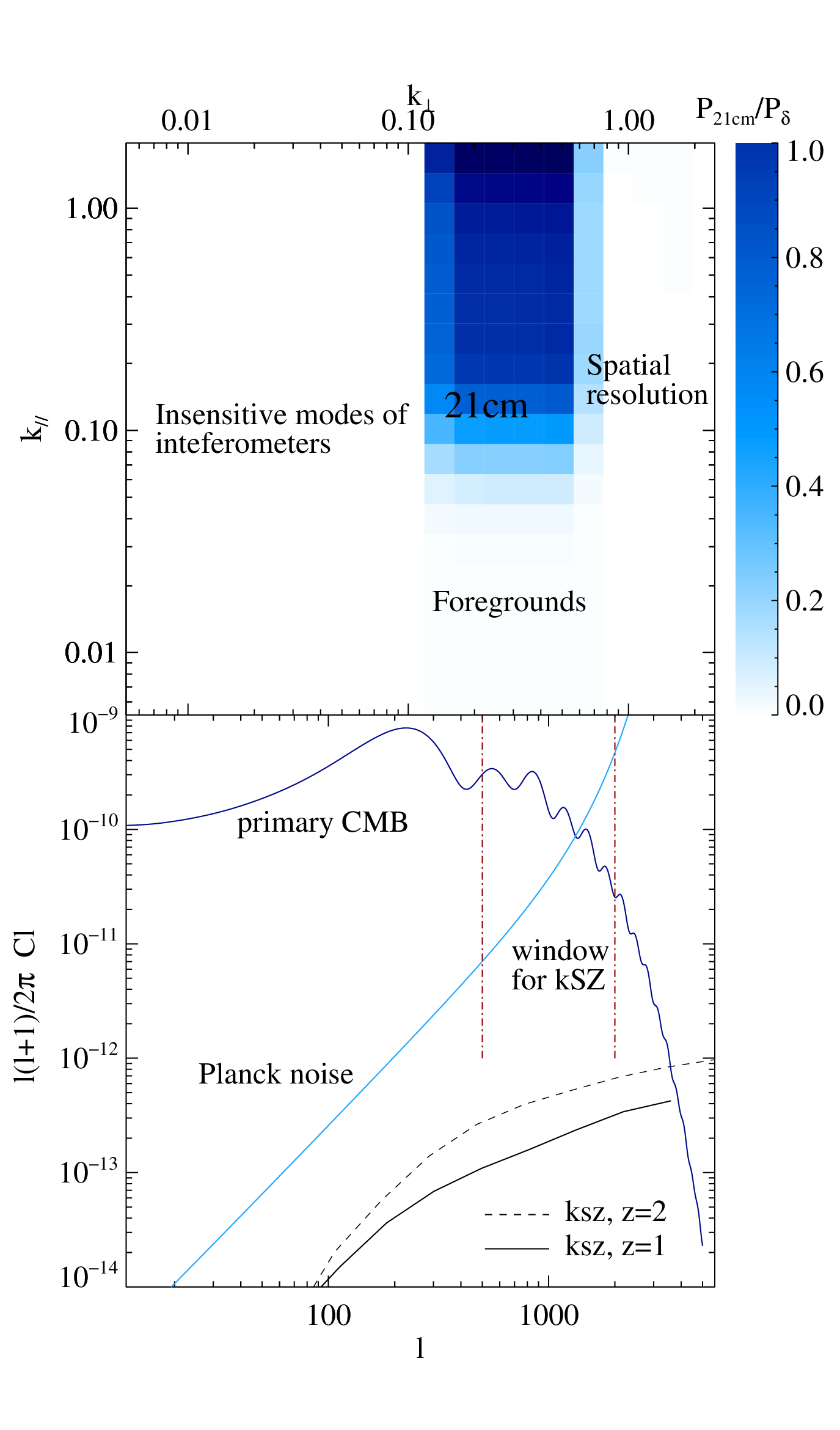}
\vspace{-0.6cm}
\end{minipage}
\begin{minipage}[t]{\linewidth}
\vspace{-0.6cm}
\center
\includegraphics[width=0.9\textwidth]{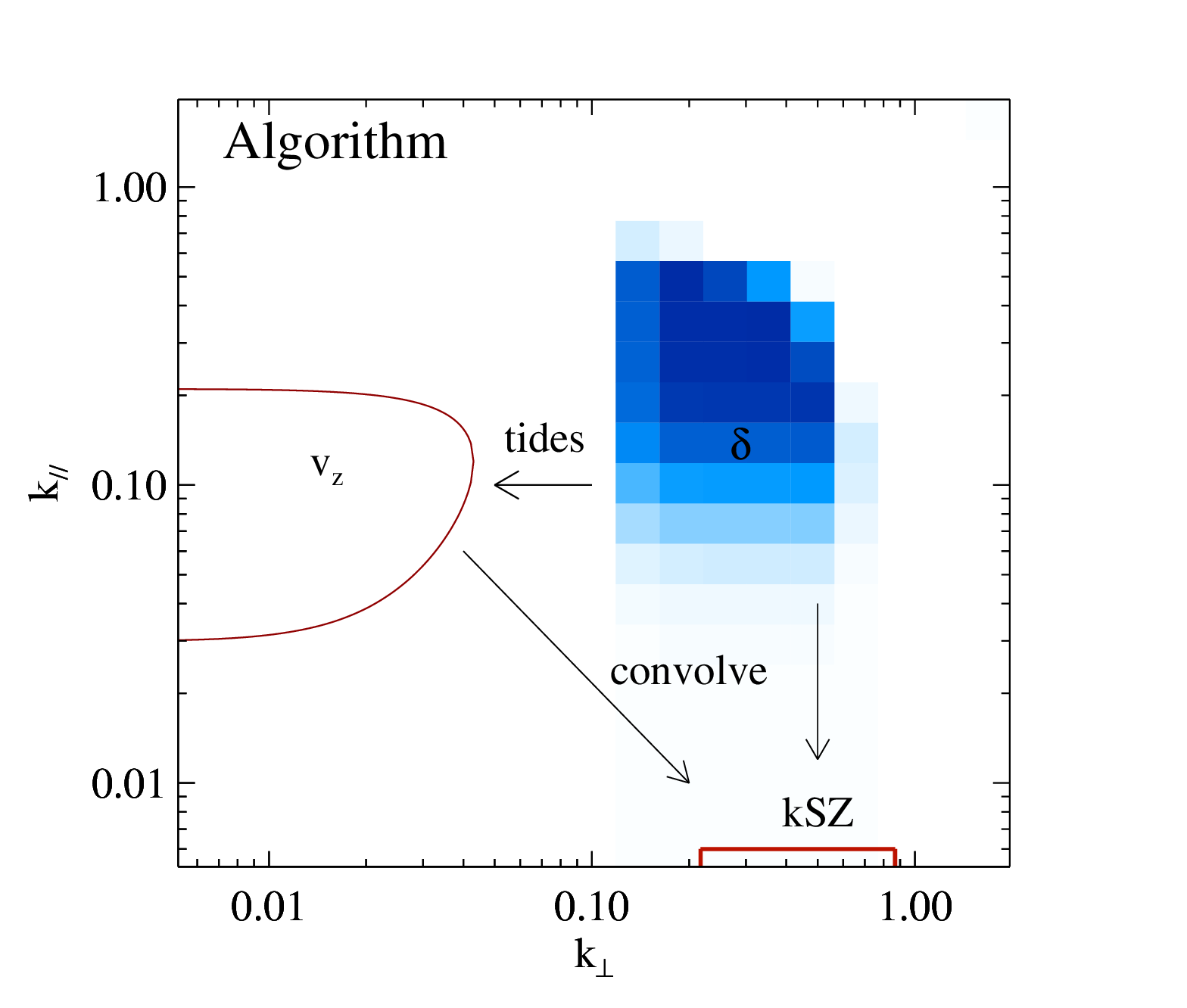}
\vspace{-0.4cm}
\end{minipage}
\caption{
({\it Top}) The detectable scales of density fluctuations in 21\,cm intensity mapping experiments with CHIME at redshift 1. 
({\it Middle}) The angular power spectrum of kSZ signals from two boxes of conformal distance 1 Gpc/$h$ centered at redshift 1 and 2 respectively, as opposed to those of primary CMB and of instrument noises from Planck at 217 GHz. 
({\it Bottom}) The generation of kSZ signals convolves density and velocity field of different spatial scales. 
We use tidal reconstruction algorithm to restore the contaminated large scale modes in intensity mapping. 
 The kSZ template is generated by convolving $v_z$ calculated from tidal reconstructed field and $\delta$ from 21\,cm intensity mapping.
}
\label{fig:cmb_21cm}
\end{figure}

\section{Cross-correlation of density fields with kSZ}
\label{sec:kszRecon}
The CMB temperature fluctuation caused by the kSZ effect is approximately a line-of-sight integral of the free electron momentum field:
\begin{eqnarray}
\label{eq:ksz}
\Theta_\mr{kSZ}(\bm{\theta})\equiv\frac{\Delta T_\mr{kSZ}}{T_{\mr{CMB}}}
=-\frac{1}{c}\int d\chi  g(\chi)  p_\parallel(\chi,\bm{\theta})\ ,
\end{eqnarray}
where $\chi(z)$ is the comoving distance, $g(\chi)=e^{-\tau} d\tau/d\eta$ is the visibility function, $\tau$ is the optical depth of Thomson scattering, 
$p_\parallel=(1+\delta_e)v_\parallel$ is the free electron momentum field parallel to the line of sight, and $\delta_e=(\rho-\bar{\rho})/\bar{\rho}$ is the free electron overdensity, with $\bar\rho$ denoting the average density. It is assumed that electron overdensity $\delta_e$ is closely related to the baryon overdensity at $z<2$, therefore, we simply use $\delta$ to denote both hereafters.

The direct correlation between kSZ and density fields vanishes due to the cancellation of positive and negative velocities, 
therefore, we follow the kSZ template method \cite{Shao11} to select for kSZ signals. 

The peculiar velocity in a radial direction could be 
calculated from the linearized continuity equation:
\begin{eqnarray}
	\label{eq:v}
v_z(\bm{k})=i a H f\delta(\bm{k})\frac{k_z}{k^2}\,
\end{eqnarray}
where $a$ is the scale factor, $f=d\ln D/d\ln a$, $D(a)$ is the linear growth function, $H$ is the Hubble parameter, 
the indice `$z$' indicates the direction 
along LOS.

We generate the kSZ template of a selected redshift bin with the measured density field $\delta$ and the calculated radial velocity field $v_z$, following Eq (\ref{eq:ksz}). 
Correlating the kSZ template with the CMB distortion selects out the kSZ signal.

To quantify the tightness of correlation between the kSZ template and actual kSZ, 
we introduce a correlation coefficient $r_\ell$:
\begin{eqnarray}
\label{eq:r}
	r_\ell\equiv \frac{C_{\ell\,\mr{tmpl,real}}}{\sqrt{C_{\ell\,\mr{tmpl}}C_{\ell\,\mr{real}}}}\,
\end{eqnarray}
where $C_{\ell\,\mr{tmpl,real}}$ is the cross angular power spectrum.

\begin{figure*}[btp]
\begin{minipage}[t]{0.325\linewidth}
\begin{center}
\includegraphics[width=\textwidth,height=1.7\textwidth]{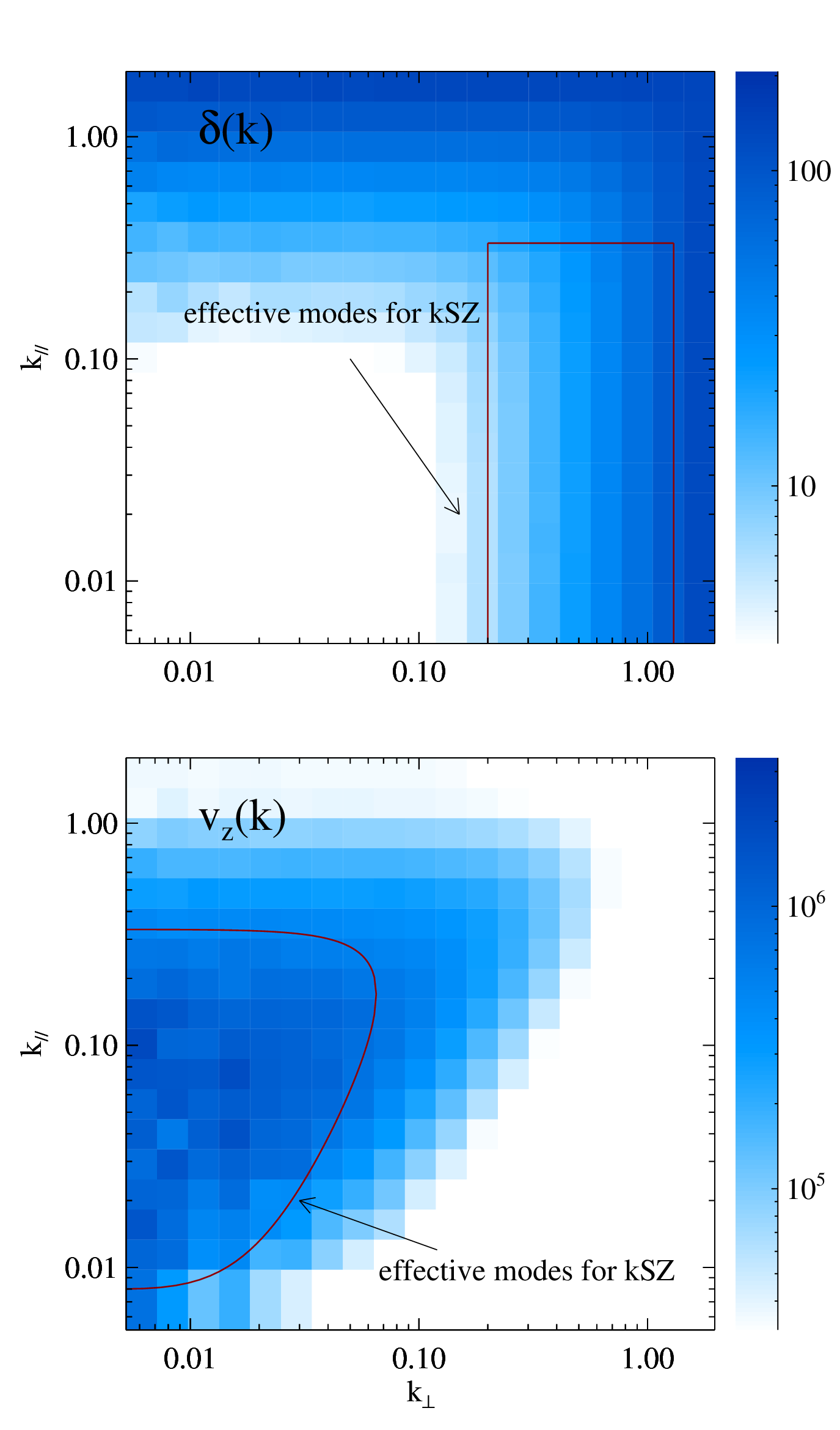}
\end{center}
\vspace{-0.7cm}
\end{minipage}
\begin{minipage}[t]{0.325\linewidth}
\begin{center}
\includegraphics[width=\textwidth,height=1.7\textwidth]{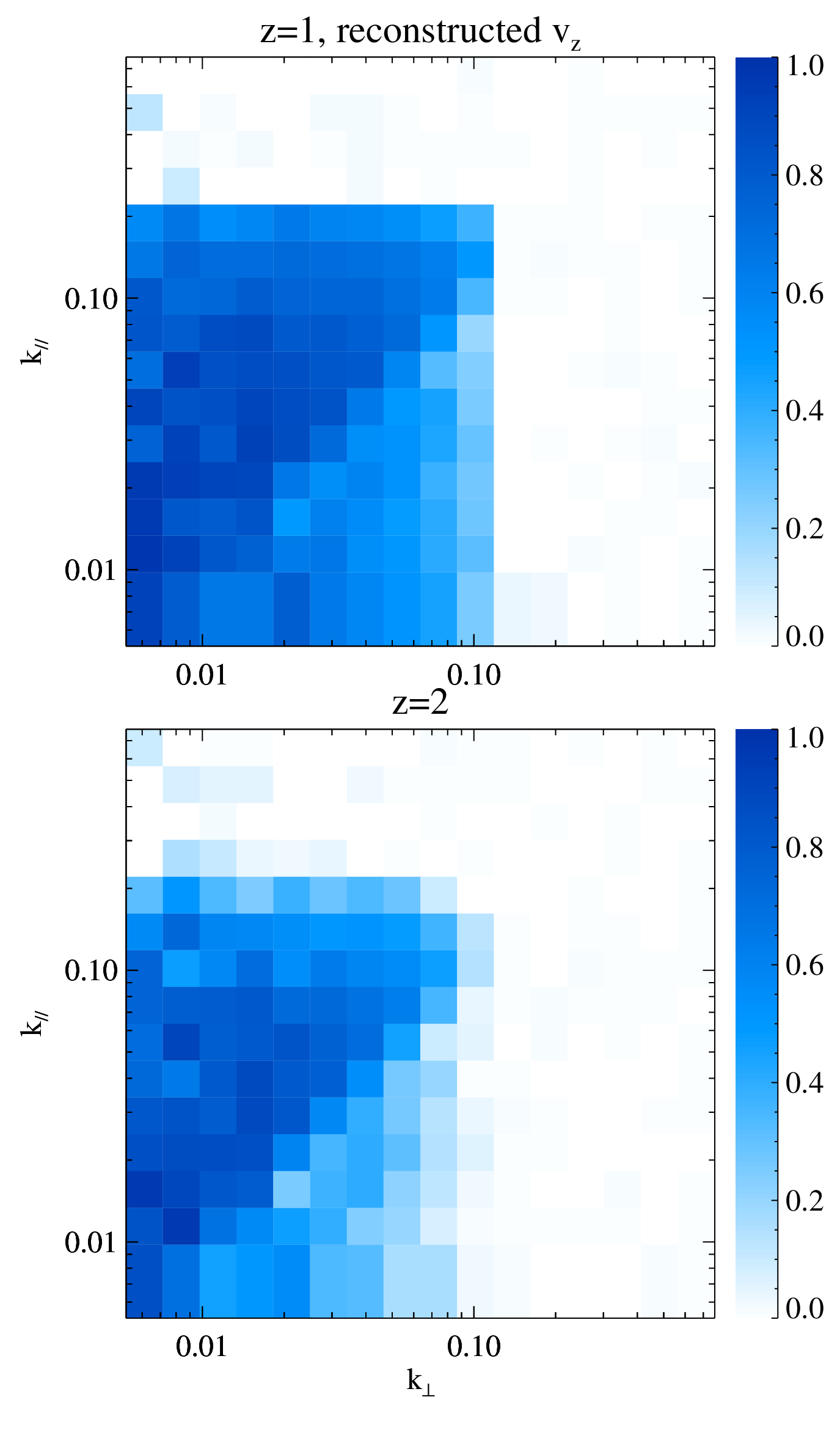}
\end{center}
\vspace{-0.7cm}
\end{minipage}
\begin{minipage}[t]{0.325\linewidth}
\begin{center}
\includegraphics[width=\textwidth,height=1.7\textwidth]{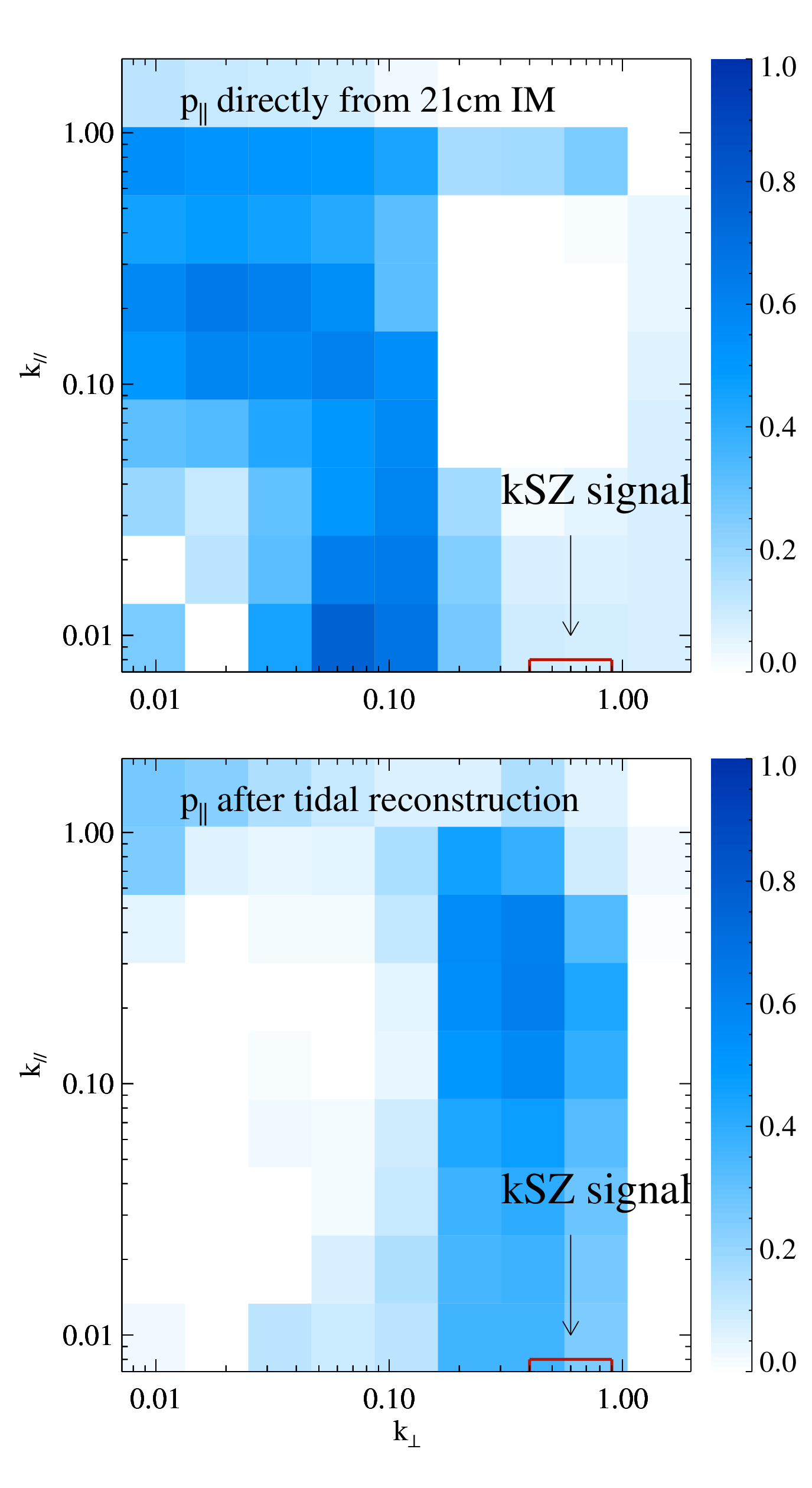}
\end{center}
\vspace{-0.7cm}
\end{minipage}
\caption{
({\it Left}) The color represents the variances $2\pi^2\Delta^2\equiv k^3P(k)$ of the complete density and velocity field at $z=1$.  
It indicates the weights of different Fourier modes contributing to fields in real space. The modes that are essential for kSZ distortions at $\ell\sim 500-2000$ are marked out. 
({\it Middle}) The correlation coefficient between the tidal reconstructed $v_z$  and actual $v_z$  at $z=1~\&~2$, 
assuming the original field is collected by CHIME with high foregrounds. 
({\it Right}) Upper panel shows the correlation coefficient between momentum fields $p_\parallel=(1+\delta)v_z$ calculated from IM fields and the actual one. 
Lower panel shows the increased correlation after tidal reconstruction. 
KSZ signal corresponds to $k_z\sim0$ modes, for which $\ell\sim  500-2000$ is marked out. 
}
\label{fig:k3v}
\end{figure*}

\section{Challenges for 21 cm intensity mapping}
\label{sec:21cm}



Given complete detection of density fluctuations, following the procedures described in the previous section, 
we should be able to retrieve $>90\%$ of the kSZ signal from CMB at selected redshift bins\,\cite{Shao11}. 
However, 21 cm IM experiments are only sensitive to density fluctuations on certain scales, because of the several sources of noises:

\begin{enumerate}
\item Foreground noises:
IM intends to use all photons to map the density field. 
While gaining unprecedented survey speed, it leads to severe foreground contamination. 
The foregrounds, typically three orders of magnitude stronger than the signals, have complicated origins, ranging from galactic emission, extragalactic radio sources, radio recombination lines to the noises from the telescopes \cite{DiMatteo04,Masui13}.  
It will contaminate the signals of large-scale structures in the radial direction. 

\item Zero spacing problem of interferometers:
Current 21cm IM experiments are all carried on interferometers ---  
on the one hand, they are stable; 
on the other hand, the cross correlations from different dishes have orders of lower noises than auto-correlations from a single dish. 
For CHIME-like facilities, with multiple beams installed on one dish, 
the calibration for cross correlation between two beams of the same dish 
are complicated. 
Therefore, we only consider signals from cross correlating different dishes for the rest of the paper. 
The minimum spacing between dishes, i.e. the shortest baseline of the interferometer, decides the largest angular scale it could probe.  
It results in an inner hole of small $k_\perp$ of sampled density field in the Fourier space. 

\item Small scale noises:
The smallest scale density fluctuations detectable in 21cm IM experiments are jointly decided by the angular resolution of the facility, i.e. the longest baseline, receiver noise and shot noise. 
For redshift one, the receiver noise dominates. It gives an upper limit of the Fourier modes we could detect.  
\end{enumerate}

Fig. \ref{fig:cmb_21cm} upper panel is an illustration of these effects in density field obtained with 21cm IM. 
Directly using it to correlate with CMB map from Planck will only retrieve $\sim10\%$ of the underlying kSZ signal.

\section{Important scales for kSZ}
\label{sec:kszAnalysis}
In this section, we discuss how different Fourier modes of density and velocity field contribute to the kSZ distortions. 

As demonstrated in Fig.\,\ref{fig:cmb_21cm}, the kSZ effect is too faint to be distinguished until the primary CMB starts to fade away, at roughly $\ell>500$.
It is possible to select a frequency band where the thermal SZ signal is negligible, then the dominant factor at high $\ell$ will be the CMB instrumental noise.  With existing Planck \cite{Planck2015} data at 217 GHz, $\ell \sim 500-2000$ will be the visible window for kSZ signal. The window could be extended to higher multipole with Simons Observatory \cite{2018SO} and CMB-S4 \cite{2016CMBS4}, but for this paper, we focus on $\ell \sim 500-2000$. 

Write the kSZ distortion $\Theta(\bm{\theta})$ from a specific redshift in Fourier space as  
\begin{align}
    \label{eq:thetak}
    \Theta(\bm{\ell}) &\propto p_\parallel(\bm{\ell},0)\\
     & \propto \int 
    d^3k\,\delta(\bm{\ell}/\chi(z)-\bm{k}_\perp,-k_\parallel) v_z(\bm{k_\perp},k_\parallel)\nonumber.
    \end{align}
where $\chi(z)$ denotes the comoving distance a redshift $z$, 
we use it to transform angular scale $\bm{\ell}$ and $\bm{k}_\perp$. Notice that $g(\chi)$ is a slow varying function, so it is treated as a  constant in this analysis for convenience.  

The convolution shows that kSZ combines density and velocity field with different angular scales and identical parallel scales, 
i.e. not $v_z(\bm{k})$ and $\delta(\bm{k})$ with identical $\bm{k}$, 
but those  
with $k_z$ of same magnitude but a seperation of $\delta \bm{k_\perp}= \ell/\chi$ being coupled together. 
Therefore, the effective modes for $v_z$ and $\delta$ come as a pair with $\bm{\ell}/\chi$ separation in Fourier space. 
A demonstration of this convolution for kSZ $\ell \sim 500-2000$ is indicated in Fig.\,\ref{fig:cmb_21cm}. 

Although the summation is over all pairs of $\delta$ and $v_z$ with a specific separation in Fourier space, 
the weight of each pair can be off by orders of magnitude depending on the power spectrum of $\delta$ and $v_z$. 
In Fig.\,\ref{fig:k3v} left panels show the variances $2\pi^2\Delta^2\equiv k^3P(k)$ of density and velocity fields at redshift 1. 
Variance is an effective way to show contribution of different scales ---\,the power spectrum $P(k)$ indicates the strength of $|\delta|$ or $|v_z|$ at scale $k$, 
and $k^3$ accounts for the integral in Eq.\,\ref{eq:thetak}, which 
is a simple estimator for the space between $k$ and $k+\Delta k$ in log-log plots. 
Combining the two plots in left panels of Fig.\,\ref{fig:k3v}, 
we show that: 
\begin{enumerate}
\item Within the range of $ 0.005\lesssim k\lesssim 2$ $h$/Mpc, 
velocity variance $2\pi^2\Delta_{v_z}^2$ is a small $k_\perp$ dominated field 
while for density variance $2\pi^2\Delta_\delta^2$, large $k_\perp$ modes play an important role. 
Therefore, all the large $\ell$ modes of kSZ comes from coupling of 
$\delta(\, \mr{large}\,k_\perp)$ with $v_z(\, \mr{small}\,k_\perp)$. 
In other words, almost same large scale modes of $v_z$ contribute to all $\ell$s of kSZ, different $\ell$ of kSZ tracks $\delta$ of different spatial scales; 
\item In the parallel direction, $k^3$ greatly down-weights the large scale $P_{v_z}$, leading to middle scale $k_\parallel$ modes contributing to  $2\pi^2\Delta_{v_z}^2$. 
Therefore, kSZ signal is not solely from large scale structures along line of sight--- a significant fraction of middle scale fluctuations up to few Mpc/h enters the signal from the convolution. 
\end{enumerate}
We mark the most relevant modes for generating kSZ signal of $\ell\sim500-2000$ with red lines in Fig.\ref{fig:k3v}. 
(Notice that although these modes contribute to $\sim90\%$ of the kSZ signals, but modes with smaller contribution can still be detectable given enough S/N, i.e. at larger $\ell$s when primary CMB fades away, it is possible to probe into even smaller scales. However, this is beyond our discussion here.)

The essential modes for density and velocity field require completely different spacial resolutions. 
Since the velocity field is linearly constructed from density field, 
an optimal survey should include essential modes of both fields. 
Comparing these essential modes with the modes resolvable in 21cm IM (shown in Fig. \ref{fig:k3v} left panels and Fig.\ref{fig:cmb_21cm} upper panel respectively), 
we notice that while the effective modes for $\delta$ are partly resolved, the large scale dominated $v_z$ is almost completely lost in the 21 cm IM.
We attempt to recover these modes with cosmic tidal reconstruction \cite{2015:zhu,2012:pen}.

\section{Cosmic tidal reconstruction}
\label{sec:tide}
The density fluctuations on different scales interact
under the gravitational interactions during nonlinear structure formation. 
The evolution of small-scale density fluctuations is modulated by the long wavelength density perturbations \cite{2015:zhu,2014Tides}. 
By studying the anisotropic tidal distortions of the local small-scale power spectrum, it is possible to solve the tidal field and hence the underlying large-scale structures \cite{2012:pen,2015:zhu,2016Zhu,2018Foreman}. 

The leading order effect of the long wavelength perturbation is described
by the large-scale tidal field,
\begin{eqnarray}
\label{eq:tij}
t_{ij}=\partial_i\partial_j\Phi_{L}-\nabla^2\Phi_L\delta_{ij}/3,
\end{eqnarray}
where $\delta_{ij}$ is the Kronecker delta function, $\Phi_L$ is the long wavelength gravitational potential.
Here we focus on the traceless tidal field since the anisotropic distortions are more robust than the change of local power spectrum amplitude which may arise due to other processes.
From Lagrangian perturbation theory, the local anisotropic matter power spectrum due to the tidal effect from large-scale density perturbation is
\begin{equation}
P(\bm{k},\tau)|_{t_{ij}}=P(k,\tau)+\hat{k}^i\hat{k}^jt^{(0)}_{ij}P(k,\tau)f(k,\tau),
\end{equation}
where  $\hat{k}^i$ is the unit vector, $P(k,\tau)$ is the isotropic linear power spectrum, the superscript $(0)$ denotes the initial time defined in perturbation calculation, and $f(k,\tau)$ is the tidal coupling function \cite{2015:zhu}. 

The tidal force tensor $t_{ij}$ is symmetric and traceless and hence can be decomposed into five independent observables:  
\begin{eqnarray}
t_{ij}=\left( \begin{array}{ccc}
\gamma_{1}-\gamma_{z} & \gamma_{2} & \gamma_{x}\\
\gamma_{2} & -\gamma_{1}-\gamma_{z} & \gamma_{y}\\
\gamma_{x} & \gamma_{y} & 2\gamma_z
\end{array} \right).
\end{eqnarray}
Therefore, from the angular dependence of the tidal shear distortions, we can solve for different components of $t_{ij}$. 
The reconstruction of gravitational tidal shear fields is described by the same 
formulation as the weak lensing reconstruction from CMB temperature fluctuations.
The tidal shear fields are given by the quadratic fields of the small-scale density
fluctuations.
As the tidal shear fields are related to second derivative of large scale gravitational potential $\Phi_L$, different components of $t_{ij}$ can be combined to get the reconstructed large-scale density field
\begin{eqnarray}
    \label{eq:largepoten}
    \delta^\mr{tide}\propto\nabla^2\Phi=\frac{3}{2}\nabla^{-2}\partial_i\partial_j t_{ij},
\end{eqnarray}
where the large-scale density information are from the convolution of small-scale
density field.

More detailed steps are described in Ref. \cite{2015:zhu}.  
We make slight adjustments as to use all 5 observables in $t_{ij}$ \cite{2018WKH} 
rather than only $\gamma_1$ and $\gamma_2$ in transverse plane. 
As for the influence of redshift distortion concerned in \cite{2015:zhu}, 
linearly, it is just a change of absolute value of $\delta(\bm{k})$ related to $k_z$, which is easy to correct.
To avoid contaminations coming from nonliear redshift distortions, 
we discard $\delta(\bm{k})$ with $k_z$ greater than a cut off scale 
$k_{z\,\mr{max}}=k_{\perp\,\mr{max}}$ when applying tidal reconstruction algorithm. 
There are no noticeable downgrading of reconstruction results on $v_z$ 
after considering redshift distortions.

\if2=1
We use all five components, since there are more intact modes in 
z direction than transverse plane, 
and $\gamma z$ is actually the most important 
shear estimater for the $k_z>k_\perp$ region 
we are most interested in.

Programming steps:\no indent

(1) Gaussianize the field, taking 
$\delta_g=\mathrm{ln}(1+\delta)$. 
This is to allieviate the problem that filter $W_i$ in Eq.(\ref{eq:wi}) heavily weights high density regions.

(2) Following gravitational lensing procedures, decompose the symmetric, traceless tidal force tensor into 5 components, 

(3) Select density distortions caused by tidal force, 
by convolving $\delta_g$ with a filter $W_i$ 
deduced from Eq.(\ref{eq:powerdistort}) 
\begin{eqnarray}
\delta^{w_i}_g(\bm{k})=W_i(\bm{k})\delta_g(\bm{k}) 
\end{eqnarray}
\begin{eqnarray}
\label{eq:wi}
W_i(\bm{k})=i \bigg[\frac{P(k)f(k)}{P_{tot}^2(k)}\bigg]^{\frac{1}{2}}\frac{k_i}{k}
=S(k)\frac{k_i}{k}\nonumber
\end{eqnarray}
$P_{tot}=P+P_{noise}$ is the observed matter powerspectrum, 
and P is theoretical matter powerspectrum,

(4) Estimate the 5 tidal tensor components from quadratic statistics.
\begin{eqnarray}
\label{eq:gamma}
\hat{\gamma}_1(\bm{x})&=&
[{\delta}^{w_1}_g(\bm{x}){\delta}^{w_1}_g(\bm{x})-
{\delta}^{w_2}_g(\bm{x}){\delta}^{w_2}_g(\bm{x})],\nonumber\\
\hat{\gamma}_2(\bm{x})&=&
[2{\delta}^{w_1}_g(\bm{x}){\delta}^{w_2}_g(\bm{x})],\nonumber\\
\hat{\gamma}_x(\bm{x})&=&
[2{\delta}^{w_1}_g(\bm{x}){\delta}^{w_3}_g(\bm{x})],\\
\hat{\gamma}_y(\bm{x})&=&
[2{\delta}^{w_2}_g(\bm{x}){\delta}^{w_3}_g(\bm{x})],\nonumber\\
\hat{\gamma}_z(\bm{x})&=&
\frac{1}{3}[(2{\delta}^{w_3}_g(\bm{x}){\delta}^{w_3}_g(\bm{x})\nonumber\\
&&-{\delta}^{w_1}_g(\bm{x}){\delta}^{w_1}_g(\bm{x})
-{\delta}^{w_2}_g(\bm{x}){\delta}^{w_2}_g(\bm{x}))],\nonumber
\end{eqnarray}

(5) Reconstruct large scale density contrast $\kappa_\mr{3D}$ from tidal tensor:
\begin{eqnarray}
\kappa_\mr{3D}(\bm{k})=\frac{1}{k^2}
&&[(k_1^2-k_2^2)\gamma_1(\bm{k})+2k_1k_2\gamma_2(\bm{k})\nonumber\\
&&+2k_1k_3\gamma_x(\bm{k})+2k_2k_3\gamma_y(\bm{k})\\\nonumber
&&+(2k_3^2-k_1^2-k_1^2)\gamma_z(\bm{k})].
\end{eqnarray}

(6) Correct bias and suppress noise with a Wiener filter.

Due to the foregrounds, the noise in $z$ direction will be different from $x$,$y$ direction, therefore we apply an anisotropic Wiener filter.
\begin{eqnarray}
	\label{eq:wiener}
    \hat \kappa_{c}(\bm{k})=\frac{\kappa_{\mathrm{3D}}(\bm{k})}{b(k_\perp,k_\parallel)}W(k_\perp,k_\parallel)\ ,
\end{eqnarray}
Bias $b(k_\perp,k_\parallel)=P_{\mathrm{k3D,}\delta}/P_\delta$ 
is the cross powerspectra between reconstructed field $\kappa\mathrm{3D}$ and original field $\delta$, 
Wiener filter $W(k_\perp,k_\parallel)=P_\delta/(P_{\mathrm{k3D}}/b^2)$.

Here $\hat \kappa_{c}$ is the output large scale density contrast we obtain from tidal reconstruction.
We use it to calculate velocity $\hat v_z^{\mathrm{tide}}$.
\fi
\section{Simulations}
\label{sec:simu}

\begin{figure}[tbp]
\begin{center}
\includegraphics[width=0.48\textwidth]{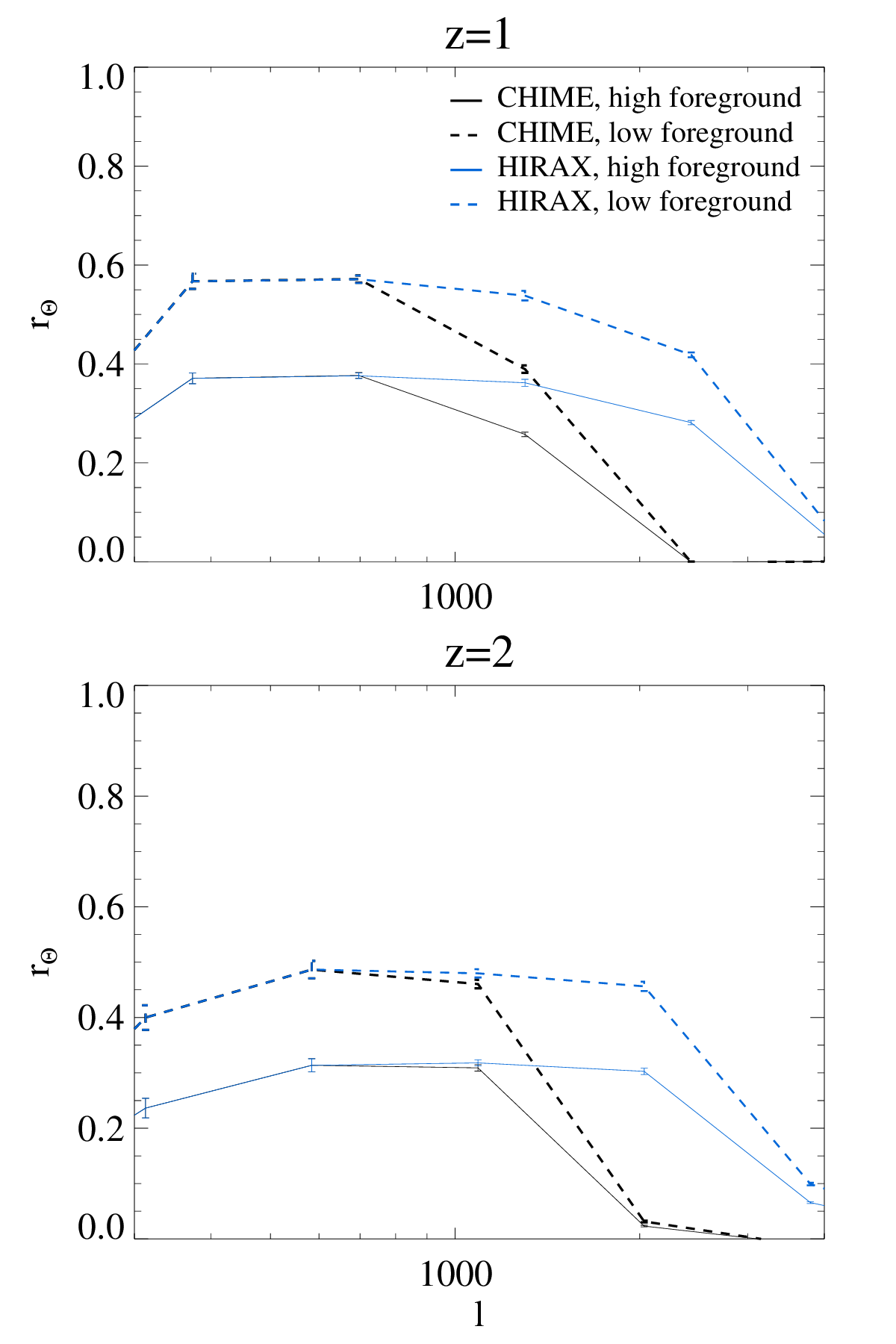}
\end{center}
\vspace{-0.7cm}
\caption{The correlation coefficient $r$ between kSZ distortions 
and kSZ templates generated from 21cm IM field. Forecasts for two instruments and different levels of foregrounds are calculated for $z=1$ (above) and $z=2$ (below). 
}
\label{fig:r}
\end{figure}

We run six $N$-body simulations, using the $\mr{CUBEP}^3\mr{M}$ code \cite{2013:code}, each evolving $1024^3$ particles in a $(1.2\, \mr{Gpc}/h)^3$ box. Simulation parameters are set as: Hubble parameter $h=0.678$, baryon density $\Omega_{b}=0.049$, dark matter density $\Omega_{c}=0.259$, amplitude of primordial curvature power spectrum $A_s=2.139\times10^{-9}$ at $k_0=0.05\;\mr{Mpc}^{-1}$ and scalar spectral index $n_s=0.968$.

\begin{table}
\begin{tabular}{|m{2cm}|m{1.5cm}|m{1.5cm}|m{1.5cm}|m{1.5cm}|}
    \hline
     & \multicolumn{2}{|c|}{$z=1$} &\multicolumn{2}{|c|}{$z=2$}\\
     \hline
     & high foreground &low foreground&high foreground& low foreground\\
     \cline{2-5}
     $R_\parallel$ Mpc/h
      & 15 & 60 & 10 & 40 \\
     \hline
     & CHIME & HIRAX & CHIME &HIRAX\\
     \cline{2-5}
     $k_{\perp\,\mr{max}}$ $h$/Mpc 
     & 0.5 & 1.2 & 0.3 & 0.8 \\
     \hline
     $\ell_\mr{min}$
     & \multicolumn{2}{|c|}{300}  & \multicolumn{2}{|c|}{200} \\
     \hline
\end{tabular}
     \caption{
        Parameters chosen to match the conditions of 21\,cm IM. 
     $R_\parallel$ indicates the largest structure resolvable along LOS after foreground removal; 
     $k_{\perp\ \mr{max}}$ represents the spatial resolution; 
     $\ell_\mr{min}$, decided by the shortest baseline, indicates the largest angular scale detectable with an interferometer.} 
     \label{tab:para}
\end{table}

We output the simulated density fields $\delta$ at $z=1\, \&\, 2$ and  
apply filters to match the conditions of real 21cm IM surveys: 
\begin{eqnarray}
\label{eq:ns}
    \delta^\mr{IM}(\bm{k})=\delta(\bm{k})H(k_{\perp\,\mr{max}}-k_\perp)W(k_\parallel)H(\ell-\ell_\mr{min})
\end{eqnarray}
where the Heaviside Function $H(k_{\perp\,\mr{max}}-k_\perp)$ describes the angular resolution of the survey:
\begin{displaymath}
H(k_{\perp\,\mr{max}}-k)=\left\{
    \begin{array}{ll}
        1,& k\leq k_{\perp\,\mr{max}}  \\
        0,& k> k_{\perp\,\mr{max}}
    \end{array}\right.,
\end{displaymath}
the high pass filter $W(k_\parallel)$ indicates the loss of information due to the foregrounds: 
\begin{displaymath}
W(k_\parallel)=1-e^{k_\parallel^2R_\parallel^2/2}\, ,
\end{displaymath}
For each redshift, we consider a high foreground case based on early observations \citep{Switzer13} and a low foreground case from theoretical predictions \citep{15Shaw}.
The other Heaviside function $H(\ell-\ell_\mr{min})$ indicates the largest angular scale detectable with an interferometer. 
\begin{displaymath}
H(\ell-\ell_\mr{min})=\left\{
    \begin{array}{ll}
        0,& \ell\leq \ell_\mr{min}  \\
        1,& \ell > \ell_\mr{min}
    \end{array}\right.,
\end{displaymath}
$\ell_\mr{min}$ is determined by the length of the shortest baseline $B_\mr{min}$, $\ell_\mr{min}\sim2\pi B_\mr{min}/(z+1)\lambda$. We conservatively choose a shortest baseline of $\sim20$\,m. 

The parameters selected for two redshifts are listed in table \ref{tab:para}.

We calculate the kSZ template following the procedure demonstrated in Fig.\ref{fig:cmb_21cm} lower panel. 
First, we solve for the missing large scale modes of $\delta_\mr{IM}$ with tidal reconstruction algorithm.  
With the reconstructed large scale density field $\delta^\mr{tide}$, we then calculate $v_z^\mr{tide}$ and cross correlate it with $\delta^\mr{IM}$ to get the kSZ template. 

Comparing the reconstructed $v_z^\mr{tide}$ field with the one directly output from simulations (Fig.\,\ref{fig:k3v} middle panels),  we can see that the large scale modes are well-reconstructed. 
For the modes that contributes heavily to kSZ distortions (see Fig.\,\ref{fig:k3v} left bottom panel), more than 70\% of the information is retrieved. 
Fig.\, \ref{fig:k3v} right panels show the increased information in the momentum field template after the tidal reconstruction. 
KSZ signal corresponds to $k_z\sim0$ modes of momentum field, 
for which 
the correlation coefficient has an obvious increase at $\ell\sim500-2000$.

The correlation coefficients between the recovered kSZ template and actual kSZ distortions from a certain redshift bin are presented in Fig.\,\ref{fig:r}. 
At $z=1$, the correlation coefficient can reach 0.6 before the small scale noises start to dominate, 
while at $z=2$, a correlation coefficient of 0.5 can be reached. 
Foreground level is still the dominate factor for the correlation coefficient. The correlation coefficient is reduced by 0.2 in the strong foreground case. 
  
\subsection{Signal to noise}
\label{sec:sn}

\begin{figure}[tbp]
\begin{center}
\includegraphics[width=0.48\textwidth]{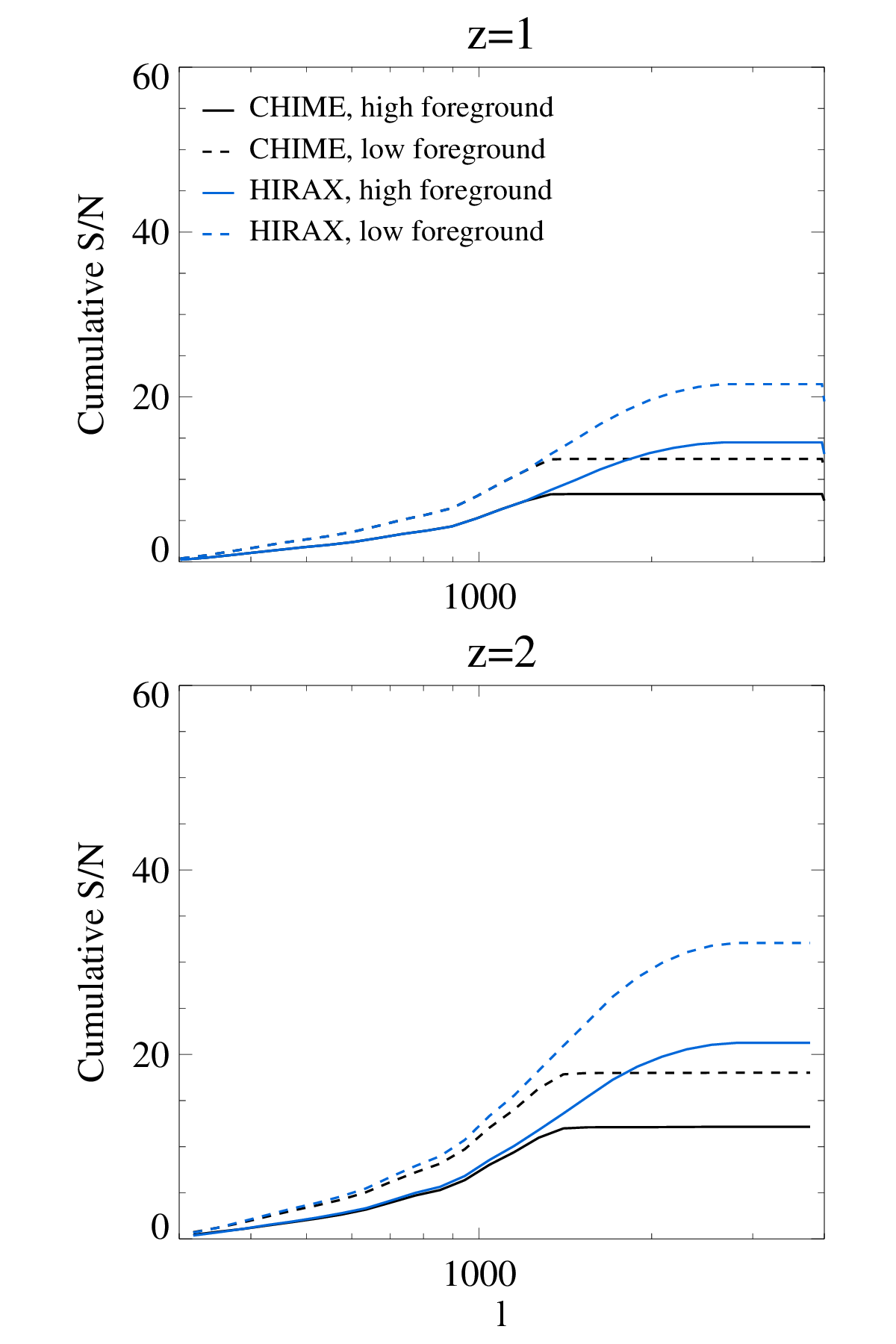}
\end{center}
\vspace{-0.7cm}
\caption{Cumulative S/N for the detection of kSZ signals by cross-correlating CMB from Planck with density fields from CHIME or HIRAX with $40\%$ sky coverage. The S/N is estimated for two redshift bins of 1 Gpc/$h$ centered at $z=1$ and 2 respectively.  
}
\label{fig:sn}
\end{figure}
The signal-to-noise for detecting kSZ effects can be estimated as \cite{Shao11}: 
\begin{eqnarray} 
\label{eq:sn}
    \frac{S}{N}&=&\frac{C_\ell}{\Delta C_\ell}\\\nonumber
               &\simeq&
    r\sqrt{(2\ell+1)\Delta l f_\mathrm{sky}}\sqrt{\frac{C_\ell^{\mathrm{kSZ},\Delta z}}{C_\ell^{\mr{CMB}}+C_\ell^\mathrm{kSZ}+C_\ell^{\mr{CMB},N}}},
\end{eqnarray}
where $C_\ell^\mathrm{CMB}$ is the angular power spectrum of primary CMB, $C_\ell^\mathrm{CMB,N}$ is the power spectrum of instrument noises, $C_\ell^{\mathrm{kSZ},\Delta z}$ is the kSZ signal from within a certain redshift bin,  
$r$ is the correlation coefficient defined in Eq.(\ref{eq:r}), and $f_\mathrm{sky}$ is the percentage of sky area covered by both CMB and 21 cm IM surveys.

We calculate $C_\ell^\mr{CMB}$ using CAMB \cite{CAMB},  and estimate $C_\ell^\mr{CMB,N}$ with Planck data \cite{Planck2015} at 217GHz. 
$C_\ell^\mr{CMB,N}=(\sigma_{p,T}\theta_\mr{FWHM})^2\,W_\ell^{-2}$, 
where 
$W_\ell=\exp[-\ell(\ell+1)/2\ell^2_\mr{beam}]$ is the smoothing window function, 
with $\ell_{beam}=\sqrt{8\ln2}/\theta_\mr{FWHM}$. 
Sensitivity per beam solid angle $\sigma_{p,T}=8.7\mu \mr{K}_\mathrm{CMB}$  
and effective beam FWHM $\theta_\mathrm{FWHM}\sim 5'$. 
 We assume sky coverage $f_\mathrm{sky}=0.4$. 
 $C_\ell^{\mathrm{kSZ},\Delta z}$ is calculated within two bins of size 1200 Mpc/h, centered at redshift 1 $\&$ 2, respectively. 

The cumulative S/N for CHIME+Planck and HIRAX+Planck is shown in Fig.\ref{fig:sn}. The S/N at $z=2$ is higher than $z=1$ due to higher electron density. 
For CHIME + Planck, the resolution of CHIME determines the largest $\ell$ detectable, 
while for HIRAX + Planck, the resolution of Planck sets the limit.
The kSZ signal is more prominent at larger $\ell$s due to the decreased strength of primary CMB, therefore 
increasing the resolution of facilities will largely increase the cumulative S/N.

\section{Ramification}
\label{sec:conclusion}
KSZ distortions come from the coupling of density and radial velocity field of different spatial scales. 
Although the final distortions on CMB are 2D, they contain information of radial structure due to the coupling of two fields.  
Different angular scale $\ell$ of kSZ tracks density field of different spatial scales, because the velocity field is large-scale dominant and contributes similarly to every $\ell$.
The strength of kSZ signals at various angular scales gives measurements of the baryon content and diffuseness. 

In this paper, we discuss the possibility of cross-correlating CMB with 21\,cm IM field as a probe for kSZ distortions. 
21\,cm IM, with fast survey speed and accurate redshift information, is promising at detecting density fluctuations for large sky area and up to high redshift. 
The biggest challenge for the cross-correlation is the loss of large-scale information in IM due to both foregrounds and interferometer zero spacing problem. 
To alleviate the problem, we reconstruct the missing large-scale modes in IM from their tidal influence on the small-scale density fluctuations. 
With $>70\%$ of the relevant large-scale information retrieved, we are able to obtain correlation coefficients for kSZ and 21\,cm fields of $0.6$ for $z=1$ and $0.5$ for $z=2$ in simulations. 
They are of obvious increase from the $\sim0.1$ correlations obtained by directly using the foreground contaminated fields. 
For $\ell\sim500-2000$, using CHIME+Planck, a detection of S/N$\sim20$ can be reached for $z\sim0.8-2.5$, while for HIRAX+Planck, a $\sim 40\sigma$ detection can be expected. 
This S/N can be greatly increased with instruments of higher spatial resolution or targeting at higher redshift where kSZ signals are more prominent, e.g., SPT-3G \cite{2014SPT3G}, Advanced ACTpol \cite{2016AdvACTpol}, Simons Observatory \cite{2018SO}, CMB-S4 \cite{2016CMBS4}, Stage II Hydrogen IM experiment \cite{2018S2},\,etc.

\if1=2
Currently the main obstacle to obtain larger correlation is the loss of $\delta( \mr{high}\, k_\perp, \mr{low}\, k_z)$ in 21\,cm IM due to foregrounds. ++++
These modes are supposed to couple with $v_z(\mr{low}\, k_z,\mr{low}\, k_\perp)$ and generate $\gtrsim30\%$ of the kSZ signals. 
The tidal reconstruction recovers large scale structures successfully, but not for recovering large $k_\perp$ modes.  
This is because the algorithm requires a greater number of known modes to solve for unknown modes. 
There are much fewer modes with small $k$ compared to the observed 21\,cm field, so they can be well-constrained. 
As $k$ gets larger, the number of modes required to solve for them increases fast and soon exceeds the number of observable modes. 
This problem can be alleviated with increasing redshift resolution --- extending the observed field to high $k_\parallel$ to gather enough information to solve for large $k_\perp$. 
In principle, 21\,cm surveys should have no problem achieving better accuracy of redshift. 
In that case, we can expect the tidal reconstruction to perform better at large $k_\perp$ and hence achieve better correlation with kSZ.
\fi

We make several approximations in the paper. We use dark matter field to assemble HI field in the analysis, which enables us to work with simple $N$ body simulations. Careful treatment should include hydrodynamic simulations which takes into account the baryonic effects (eg.\,\cite{2018HIsimulation}). 
We ignore the foreground wedge, because it is not an intrinsic loss of information, and is believed to be removable with better understanding of instruments \cite{16Liu}.
We use two boxes of 1.2\,Gpc/h width output at $z=1\,\&\,2$ to represent the density field of $z=0.8-2.5$. This ignores the redshift evolution within each box. More careful work should include thinner boxes output at different redshifts. 
We use a uniform weight when summing over different redshifts and angular scales. The S/N could be improved if proper weights are assigned \cite{2018KSZ1}. 
However, the current setups are good enough to demonstrate the feasibility of cross correlating CMB and 21\,cm IM after tidal reconstruction to probe kSZ signals. 
 
Cross-correlating the kSZ signal with 21 cm IM is promising due to its feasibility with near-term data. CHIME has started collecting data, and construction for HIRAX is underway. It is reasonable to expect our method to be testable within the next five years. This may foster understanding of stellar feedback at the scale of galaxy clusters and filaments and therefore the evolution of large-scale structures.

\section*{Acknowledgements}
We appreciate the comments from Philippe Berger on the manuscript.
The simulations are performed on the BGQ supercomputer 
at the SciNet HPC Consortium.
SciNet is funded by: the Canada Foundation for Innovation under the auspices 
of Compute Canada;
the Government of Ontario; the Ontario Research Fund -- Research Excellence;
and the University of Toronto.
Research at the Perimeter Institute is supported by the Government of Canada through Industry Canada and by the 
Province of Ontario through the Ministry of Research $\&$ Innovation. 
The Dunlap Institute is funded through an endowment
established by the David Dunlap family and the University of Toronto.

\bibliographystyle{apsrev}
\bibliography{ksz}
\end{document}